\documentclass[english,twocolumn,superscriptaddress]{revtex4-1}
\usepackage[T1]{fontenc}
\usepackage[utf8]{inputenc}
\usepackage{graphicx}
\usepackage{amssymb}
\usepackage{amsmath}
\usepackage{amsfonts,epsfig,graphicx,latexsym,bm,color,braket}
\usepackage{siunitx}

\usepackage{epstopdf}	

\usepackage{xcolor}

\allowdisplaybreaks[2]
\usepackage{babel}

\begin{document}
\title{Quantum degenerate Fermi gas in an orbital optical lattice}

\author{M. Hachmann}
\affiliation{Institut f\"ur Laserphysik, Universit\"at Hamburg, 22761 Hamburg, Germany}
\affiliation{Zentrum f\"ur Optische Quantentechnologien, Universit\"at Hamburg, 22761 Hamburg, Germany}
\author{Y. Kiefer}
\affiliation{Institut f\"ur Laserphysik, Universit\"at Hamburg, 22761 Hamburg, Germany}
\affiliation{Zentrum f\"ur Optische Quantentechnologien, Universit\"at Hamburg, 22761 Hamburg, Germany}
\author{J. Riebesehl}
\affiliation{Institut f\"ur Laserphysik, Universit\"at Hamburg, 22761 Hamburg, Germany}
\author{R. Eichberger}
\affiliation{Institut f\"ur Laserphysik, Universit\"at Hamburg, 22761 Hamburg, Germany}
\affiliation{Zentrum f\"ur Optische Quantentechnologien, Universit\"at Hamburg, 22761 Hamburg, Germany}
\author{A. Hemmerich}
\affiliation{Institut f\"ur Laserphysik, Universit\"at Hamburg, 22761 Hamburg, Germany}
\affiliation{Zentrum f\"ur Optische Quantentechnologien, Universit\"at Hamburg, 22761 Hamburg, Germany}
\affiliation{The Hamburg Center for Ultrafast Imaging, Universit\"at Hamburg, 22761 Hamburg, Germany}

\begin{abstract}
Spin-polarized samples and spin mixtures of quantum degenerate fermionic atoms are prepared in selected excited Bloch bands of an optical chequerboard square lattice. For the spin-polarized case, extreme band lifetimes above $10\,$s are observed, reflecting the suppression of collisions by Pauli's exclusion principle. For spin mixtures, lifetimes are reduced by an order of magnitude by two-body collisions between different spin components, but still remarkably large values of about one second are found. By analyzing momentum spectra, we can directly observe the orbital character of the optical lattice. The observations demonstrated here form the basis for exploring the physics of Fermi gases with two paired spin components in orbital optical lattices, including the regime of unitarity.
\end{abstract}

\maketitle
Optical lattices are synthetic arrays of bosonic or fermionic neutral atoms or molecules trapped in laser-induced periodic potentials \cite{Gry:01}. Aside from their practical use in atomic clock applications \cite{Lud:15} they are celebrated as an ideal toolbox for quantum simulation of lattice physics \cite{Fey:82, Jak:98, Lew:07, Gro:17}. Their usefulness in the context of quantum simulation of electronic crystalline matter requires in particular the use of fermionic particles, which assume the role of the electrons tunneling and interacting in a lattice of ionic cores. In fact, there is a promising strain of research devoted to emulate the fermionic Hubbard model \cite{Hub:63} and to experimentally explore its phase diagram \cite{Joe:08, Sch:08, Har:15, Gre:16, Bro:17, Maz:17, Tar:18}, which on the theory side even with modern computational power has remained an open challenge. However, many of the intriguing functionalities of crystalline electronic condensed matter rely on orbital degrees of freedom, which play a decisive role for metal-insulator transitions, superconductivity and colossal magnetoresistance in transition-metal oxides \cite{Tok:00, Mae:04}. Orbital $p$-like single-particle wave functions have been recently simulated with electrons in the second band of an artificial square lattice formed by an array of carbon monoxide molecules on a Cu(111) surface \cite{Slo:19}. It is however not obvious, how this scenario could be extended to emulate many-body physics. A natural but insufficient approach to extend optical lattices with fermionic atoms to include higher Bloch bands, is to load sufficiently many atoms \cite{Blo:08}. This, however, requires multiply occupied lattice sites and hence leads to deleterious collisions of more than two particles resulting in excessive loss and heating in connection with molecule formation \cite{Wei:03, Chi:10}.

\begin{figure}[h]
\includegraphics[width=8.2 cm]{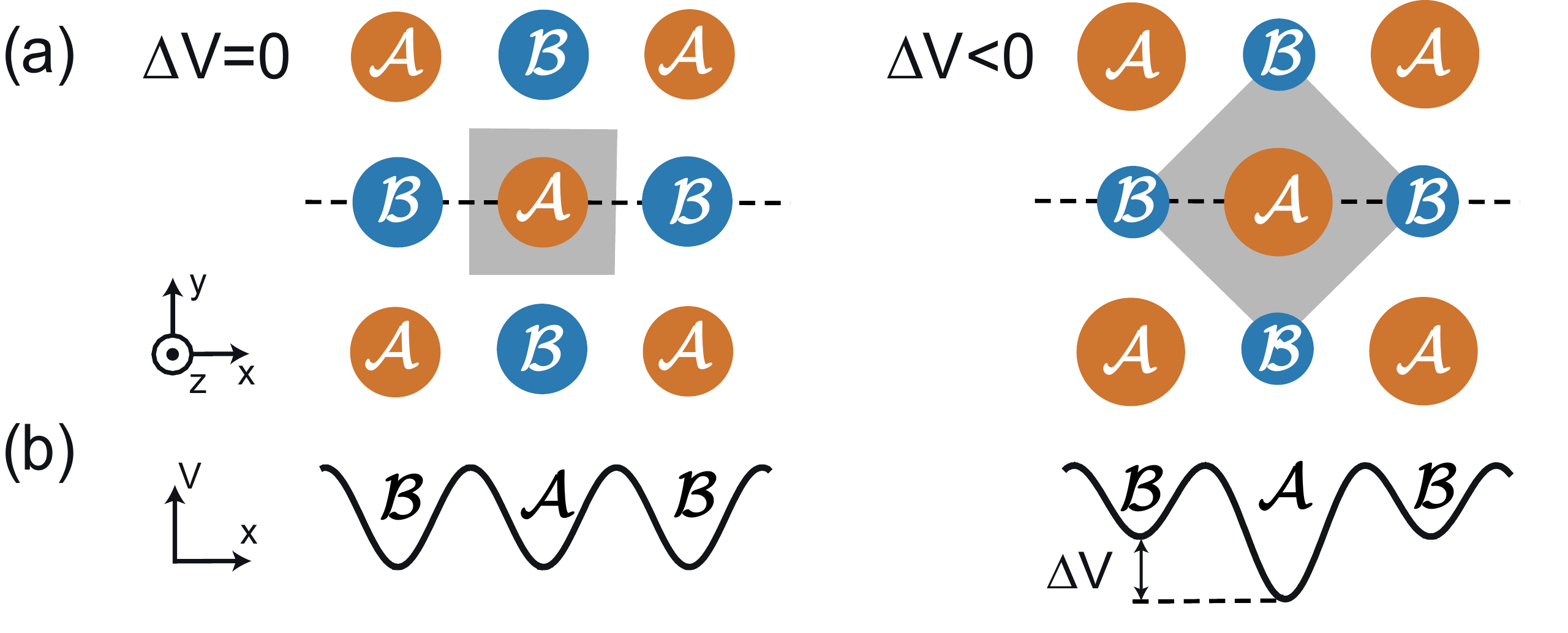}
\caption{(a) Lattice geometry in the $xy$-plane for $\Delta V = 0$ (left panel) and $\Delta V < 0$ (right panel). The grey squares denote the unit cell of the lattice. (b) Sections of the lattice potential along the dashed lines in (a).}
\label{fig1}
\end{figure}

An alternative approach, that was pioneered for bosonic atoms, selectively excites the atoms from the lowest band into a desired higher target band, thus keeping the site occupation low \cite{Wir:11, Lew:11, Oel:11, Koc:16}. The underlying strategy is that the functionality of interest takes place in a higher band and does not discriminate between a filled or an empty lowest band. There is reason to assume that two-body collisions would lead to immediate band relaxation. However, theoretical \cite{Sto:08, Pau:13} and experimental \cite{Wir:11, Koc:16} research has shown, that with appropriately designed lattice geometries reasonably long lifetimes can be realized, which has triggered widespread interest in optical lattices with orbital character \cite{Li:16}. 

\begin{figure}[ht]
\includegraphics[width=9.2 cm]{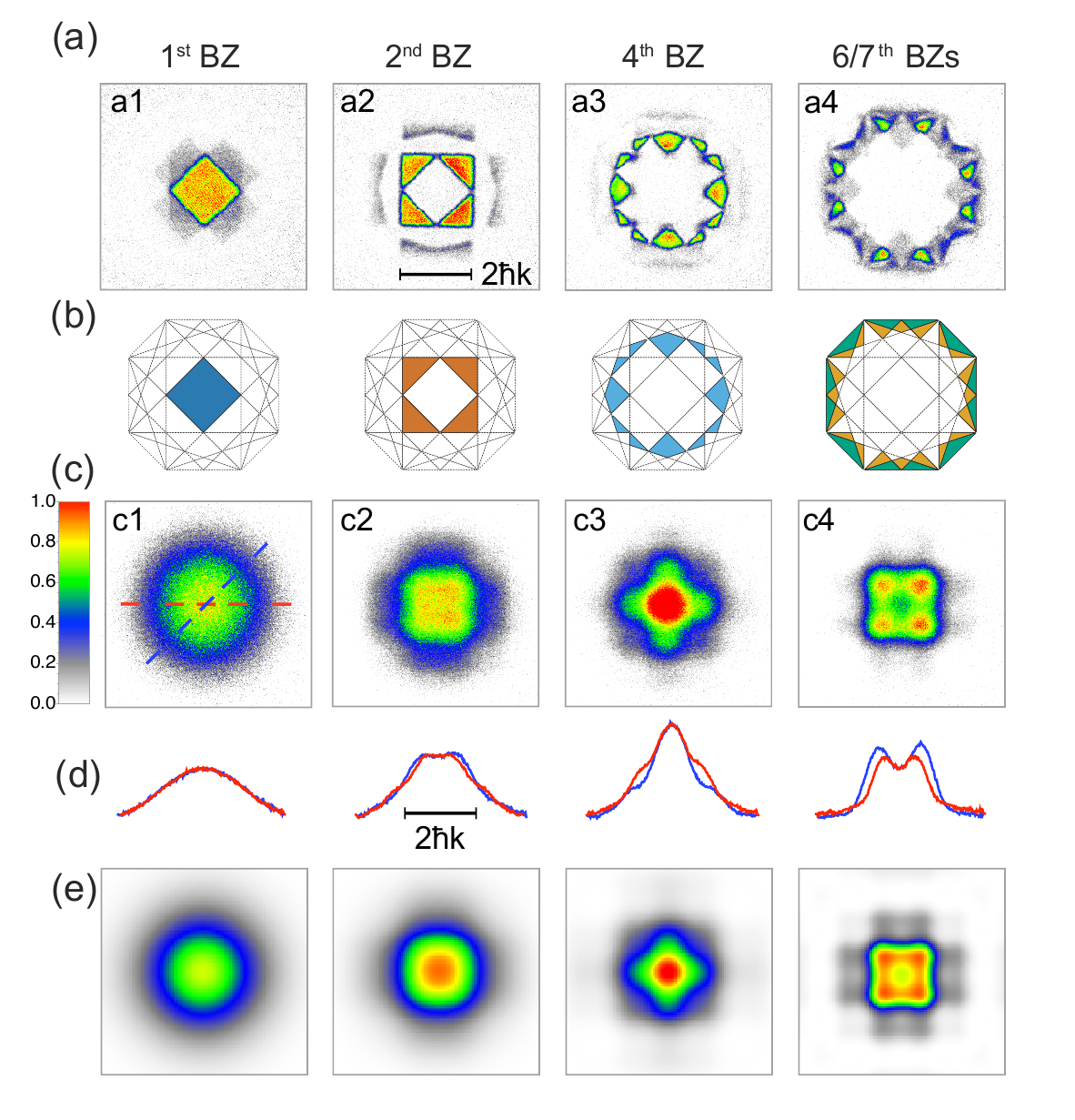}
\caption{(a) Band mapping images showing the population of the $n^{\textrm{th}}$ band in the $n^{\textrm{th}}$ Brillouin zones (BZs). Panel (a1) shows the case, if no excitation is applied, the other panels (a2), (a3), (a4) show the cases of excitation to the $2^{\textrm{nd}}$, $4^{\textrm{th}}$ and $7^{\textrm{th}}$ bands, respectively. (b) The map of BZs is shown with the first (panel 1), second (panel 2), 4th (panel 3) and 6th and 7th (panel 4) BZs highlighted. (c) Regular momentum spectra corresponding to the band mapping images in (a). The colour map on the left edge shows the normalized optical density. (d) Horizontal (red line graphs) and diagonal (blue line graphs) sections through the momentum spectra in (c) along the red dashed and blue dashed lines in (c1), respectively. (e) Momentum spectra for completely filled $1^{\textrm{st}}$, $2^{\textrm{nd}}$, $4^{\textrm{th}}$ and $7^{\textrm{th}}$ bands (from left to right) calculated for the same parameters as used in (c). The particle numbers in (c) and (e) are normalized to unity and parametrized with the same color map, shown in (c). In all plots in (a,c,d,e) $V_0 = 12 \,E_{\textrm{rec}}$.}
\label{fig2}
\end{figure}

For the first time, similar techniques are used in this work to form fermionic optical lattices with orbital degrees of freedom, which should prove useful as an advanced generation of quantum simulators for electronic matter beyond $s$-band lattice physics. For spin-polarized samples and mixtures of two spin components, the efficiency, with which selected excited bands can be occupied, as well as the corresponding lifetimes are shown to notably exceed the previous findings for bosons. We present exemplary results on the loading efficiency for the $2^{\textrm{nd}}$, $4^{\textrm{th}}$ and $7^{\textrm{th}}$ bands, but also higher bands can be addressed. For spin-polarized samples, we observe lifetimes above $10\,$s, limited by technical heating processes. Binary collisions, expected to be suppressed by Pauli's principle, are observed to play no role in this case. In contrast, for spin mixtures, two-body collisions between different spin components are observed to reduce the lifetimes. However, reasonably large values on the order of a second are also found in this case. Momentum spectroscopy confirms the orbital character of the formed wave functions. The techniques shown here form the basis for exploring the physics of Fermi gases with two paired spin components in orbital optical lattices, including the regime of unitarity, and hence may provide new fundamental insights into fermionic superfluidity in presence of orbital degrees of freedom \cite{Gre:03, Joc:03, Reg:04, Bar:04, Zwi:04, Bou:04, Par:05, Ran:12}. 

As the initial step in our experiments, a spin-polarized degenerate Fermi gas of up to $\SI{2.5 E5}{}$ potassium atoms ($^{40}$K) in the hyperfine state $\left|F=9/2, m_F=9/2\right\rangle$ with a temperature $T = 0.18 \,T_{F}$ is produced in an optical dipole trap, formed by two crossed laser beams with a wavelength of $1064\,$nm. Radio-frequency techniques can be optionally applied to prepare balanced spin mixtures of $\left|F=9/2, m_F=-9/2\right\rangle$ (spin-up) and $\left|F=9/2, m_F=-7/2\right\rangle$ (spin-down) atoms (see Ref.~\cite{SupMat} for details). The atoms are adiabatically loaded into a bipartite optical square lattice, formed by two mutually orthogonal optical standing waves with the same wavelength $\lambda = 1064\,$nm and aligned along the $x$- and $y$-axes, respectively. The optical standing waves are formed in a Michelson-Sagnac interferometer, that provides precision control of the associated band structure (see Ref.~\cite{SupMat} for details). The resulting lattice potential is composed of deep and shallow potential wells arranged as the black and white squares of a chequerboard, denoted $\mathcal{A}$ and $\mathcal{B}$, respectively \cite{Koc:16}. In the $xy$-plane, the lattice potential is approximated by
\begin{eqnarray}
V(x,y)=&-&V_{0}\left(\cos^{2}(kx)+\cos^{2}(ky)\right)\nonumber\\
&-&\frac{1}{2}\,\Delta V\cos(kx)\cos(ky)
\label{eq:potential}
\end{eqnarray}
with the wave number $k=\frac{2\pi}{\lambda}$.
Along the $z$-direction the atoms are weakly confined by an approximately harmonic potential, such that the lattice wells acquire a tubular shape. The potential depth $V_{0} \geq 0$ and the potential difference between $\mathcal{A}$-wells and $\mathcal{B}$-wells $\Delta V \in V_{0} \times [-4,4]$, can be controlled much faster than all relevant dynamical time scales. The lattice geometry in the $xy$-plane is sketched in Fig.~\ref{fig1}(a) for $\Delta V = 0$ and $\Delta V < 0$ in the left and right panels, respectively. In (b), sections through the lattice potential along the dashed lines in (a) are shown. For $\Delta V = 0$, a monopartite lattice (i.e. with equal $\mathcal{A}$- and $\mathcal{B}$-wells) is formed. Negative $\Delta V$ indicates deep $\mathcal{A}$-wells and shallow $\mathcal{B}$-wells and vice versa for positive $\Delta V$. 

After the atoms (spin-polarized or spin mixtures) are loaded to the lowest Bloch band of the optical lattice by slowly ramping up $V_{0}$ from zero to $5 -15\,E_{\textrm{rec}}$ in $150\,$ms, a quench protocol similar to that previously applied to bosonic atoms, is used to transfer them into a selected higher Bloch band. Here, $E_{\textrm{rec}} \equiv \hbar^2 k^2 / 2 m$ denotes the single-photon recoil energy and $m$ the atomic mass. The central step is to rapidly tune $\Delta V$ from negative to positive values in typically $100\,\mu$s. This technique has been summarized for bosons in Ref.~\cite{Koc:16} and a more detailed explanation adapted to the present work with fermions is provided in Ref.~\cite{SupMat}. The populations of the Bloch bands are observed by means of a standard technique referred to as band mapping (cf. Ref.~\cite{SupMat}). 

\begin{figure}[ht]
\includegraphics[width=8.8 cm]{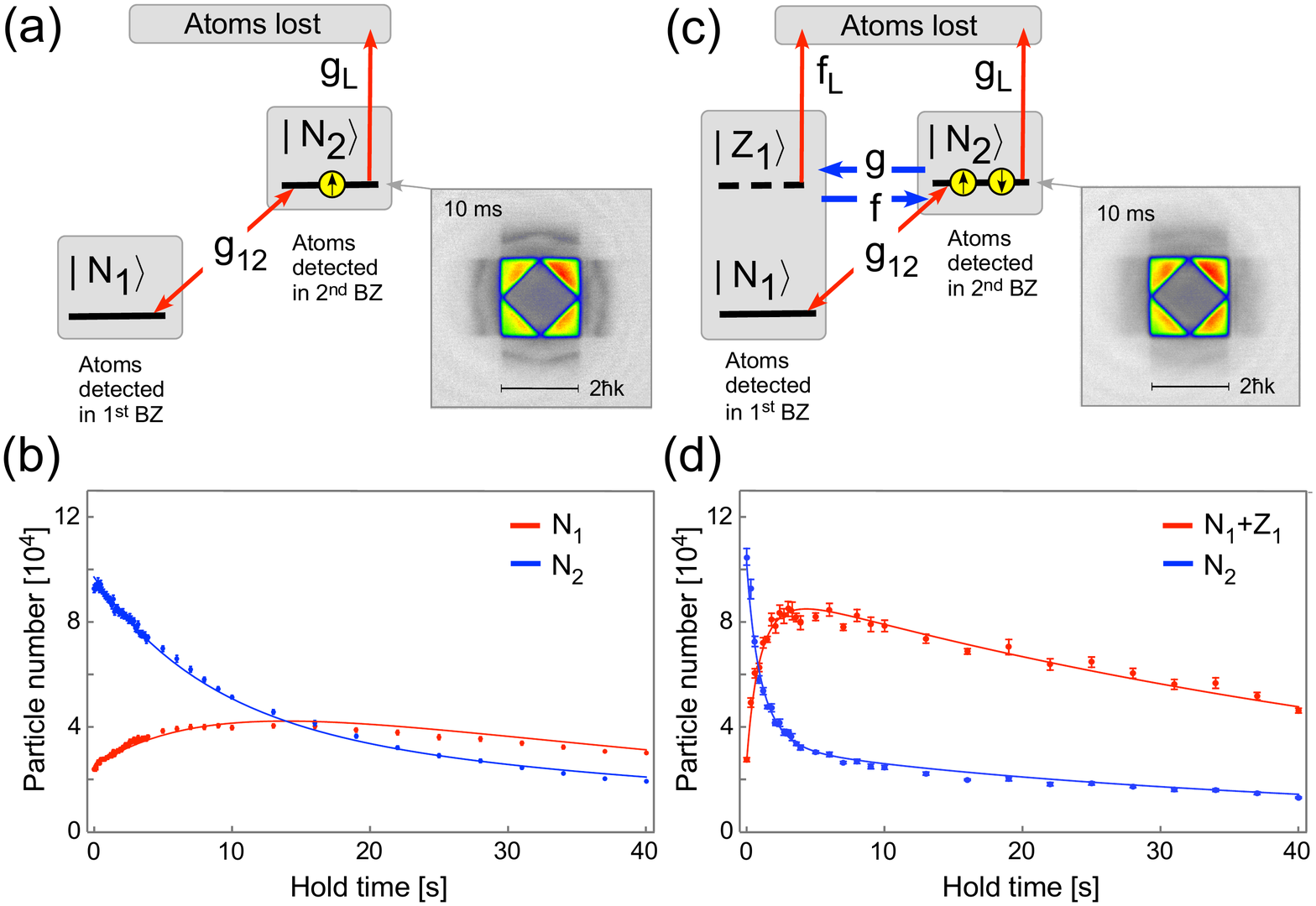}
\caption{(a) Model for band relaxation dynamics of spin-polarized samples after the $2^{\textrm{nd}}$ band is selectively populated, according to the band mapping image showing dominant population of the $2^{\textrm{nd}}$ BZ. (b) Observed populations in the $1^{\textrm{st}}$ ($N_1$: red symbols) and $2^{\textrm{nd}}$ ($N_2$: blue symbols) bands are plotted versus the holding time. The solid traces result from the fit model in (a). (c) Extended relaxation model for balanced spin mixtures, including a class of atoms with population $Z_1$ residing in the $1^{\textrm{st}}$ band with additional excitation of motion along the $z$-axis. Binary collision processes (illustrated by blue arrows) exchange pairs of spin-up and spin-down particles between $|N_2\rangle$ and $|Z_1\rangle$. (d) Observed populations in the $1^{\textrm{st}}$ ($N_1+Z_1$: red symbols) and $2^{\textrm{nd}}$ ($N_2$: blue symbols) bands are plotted versus the holding time. The solid traces result from the fit model in (c). In (b) and (d), $V_0 = 7\,E_{\textrm{rec}}$ and $\Delta V = 3.1\,E_{\textrm{rec}}$. The error bars in (b) and (d) show statistical errors for averages of $20-30$ experimental runs.}
\label{fig3}
\end{figure}

In Fig.~\ref{fig2}(a) band mapping images for spin-polarized samples are shown without excitation to higher bands (panel (a1)) and after the excitation protocol is applied to selectively excite the atoms to the $2^{\textrm{nd}}$, $4^{\textrm{th}}$, and $7^{\textrm{th}}$ band (panels (a2), (a3), (a4)). These images were recorded after the atoms were held in the lattice for $50\,$ms with $V_0 = 12 \,E_{\textrm{rec}}$ and $\Delta V_{f}/V_0 \in \{-1.24, 0.314, 0.995, 1.703\}$. The choices of $V_{f}$ adjusted for populating the $2^{\textrm{nd}}$, $4^{\textrm{th}}$, and $7^{\textrm{th}}$ bands, according to an exact band calculation, provide optimal selectivity since they maximize the gaps between the target band and adjacent bands (cf. Ref.~\cite{SupMat}). A comparison with the theoretically expected Brillouin zones (BZs) in Fig.~\ref{fig2}(b) shows that in panels (a2) and (a3), the $2^{\textrm{nd}}$ and $4^{\textrm{th}}$ BZs are selectively populated, respectively, with remarkable efficiency. In panel (a4), the 6th and 7th BZ shows population in accordance with the expectation of a band crossing between the 6th and 7th band occurring during the band mapping procedure, as predicted by an exact band calculation (cf. Ref.~\cite{SupMat}). The total fractions of atoms prepared in the 1st, $2^{\textrm{nd}}$, $4^{\textrm{th}}$, $7^{\textrm{th}}$ bands, normalized to the total number of atoms initially loaded into the lattice, are $0.67$, $0.67$, $0.57$, $0.62$, respectively. Note that due to quantum pressure of the fermionic atoms, finite temperature, and the trap potential, without excitation, only $2/3$ of the atoms are prepared in the first band, while the rest is found in higher bands (c.f. Fig.~\ref{fig2}(a1)). If we account for this circumstance and normalize the number of particles in the target bands after excitation by the number of atoms loaded to the first band, if no excitation is applied, one obtains remarkable fractions of  $0.996$, $0.84$, and $0.92$ for population of the $2^{\textrm{nd}}$, $4^{\textrm{th}}$, $7^{\textrm{th}}$ bands, respectively. Very similar results are found for spin mixtures.

In Fig.~\ref{fig2}(c), we show regular momentum spectra (cf. Ref.~\cite{SupMat}), recorded after the atoms have dwelled for $50\,$ms in the lattice, which exhibit direct signatures of the orbital character of the optical lattices formed in the $2^{\textrm{nd}}$, $4^{\textrm{th}}$, and $7^{\textrm{th}}$ bands. The shown images directly correspond to the band mapping images in Fig.~\ref{fig2}(a). These momentum spectra are expected to display squared absolute values of the Fourier transforms of the prevailing Wannier functions. For the case of panel (c1), the atoms reside in the local $s$-orbitals of the lowest band, in accordance with the observation of a perfectly isotropic momentum distribution. In panel (c2), the second band is populated and hence the atoms populate both $s$-orbitals in the shallow wells and $p$-orbitals in the deep wells. In fact, the momentum distribution appears as a superposition of a large $s$-like component as in (c1), however less localized, and a small $p$-like component that displays a cloverleaf structure with an extra node in the center. This is better seen in the sections through the images in (c1 - c4) shown in (d). The red (blue) line graphs show sections along the red dashed horizontal (blue dashed diagonal) line indicated in (c1). The superposition of $s$- and $p$-contributions explains the nearly flat top seen in the sections below (c2). In (e), calculated momentum spectra are shown, which reproduce the main features of the observations in (c1 - c4). The images in (e) result from an exact band calculation for the lattice parameters applied in (c1 - c4), neglecting the finite system size, the effect of the trap potential, and assuming that exclusively the target band is completely filled. 

The lifetime for bosonic quantum gases in higher bands is limited by two-body $s$-wave collisions \cite{Liu:06, Pau:13, Li:16, Koc:16}. In Ref.~\cite{Nus:20}, it has been shown that specific parameter configurations can be found, where different scattering processes destructively interfere with the result of remarkably long lifetimes on the order of several $100\,$ms. In the following, we explore the band decay dynamics after exciting a large fraction of fermionic atoms to the $2^{\textrm{nd}}$ band. In the case of spin-polarized samples, $s$-wave-scattering is suppressed by Pauli's principle, and the first higher order scattering contribution, i.e., $p$-wave scattering, is negligible at the given low temperatures well below $100\,$nK. At the same time, collisions with hot background atoms are negligible on the few ten second timescale, investigated here, as confirmed by the observation of lifetimes in the dipole trap of several minutes. Hence, interaction is expected to be practically irrelevant for band relaxation of spin-polarized samples. This gives rise to extreme band lifetimes, which are about two orders of magnitude longer than what has been observed with bosons. The main limitation is expected to arise through heating processes due to shaking of the lattice potential resonant with interband transitions. Heating with respect to the $z$-direction, confined by a weak harmonic potential, is expected to be comparatively small. For a minimal model of the band decay dynamics, we consider the populations of the first and second bands, $N_1$ and $N_2$, respectively, and in addition the population $N_L$ of all other bands, that are assumed not to be confined by the lattice potential and are hence considered as lost from the system. Heating couples the populations $N_1$ and $N_2$ by balanced transfer rates $g_{12}$. In addition, $N_2$ loses atoms towards $N_L$ at a rate $g_L$, which gives rise to the two equations 
\begin{eqnarray}
\label{eq:relax1}
\dot{N_1} &=& g_{12} (N_2 -N_1) \, , \\
\dot{N_2} &=& g_{12} ( N_1 - N_2) - g_L N_2 \, . \nonumber 
\end{eqnarray}
This minimal rate equation model is illustrated in Fig.~\ref{fig3}(a). At $t=0$ about $10^5$ spin-polarized fermions are prepared in the second band such that $N_2(0)/N_1(0)\approx 5$. The band mapping image in Fig.~\ref{fig3}(a) shows the initial distribution of atoms across the BZs at $t=10\,$ms, confirming predominant occupation of the $2^{\textrm{nd}}$ BZ. The observed time evolution of $N_1(t)$ (red symbols) and $N_2(t)$ (blue symbols) is shown in Fig.~\ref{fig3}(b). The model in (a) is used to determine the parameters $g_{12} =0.0456\pm 0.0015$ and $g_{L}=0.0432\pm 0.0009$ by simultaneously fitting with respect to both data sets in (b). An analytic solution of this model shows that the decay of $N_2(t)$ is exponential during the first $10\,$s with a $1/e$ decay time of $16.1\,$s.

For mixtures of the two spin components $\left|m_F=-9/2\right\rangle$ and $\left|m_F=-7/2\right\rangle$, $s$-wave collisions between different spin states are possible. The singlet and triplet scattering lengths at zero magnetic field are $105\,a_0$ and $176\,a_0$, respectively ($a_0=$ Bohr radius) \cite{Boh:00}. For modeling band relaxation, only band-index changing collisions are relevant. We assume that, similarly as found for bosons in the same lattice potential \cite{Pau:13, Nus:20}, the dominant collisional process leading to loss of $2^{\textrm{nd}}$ band population, is associated with a transfer of pairs of colliding spin-up and a spin-down atoms to the $1^{\textrm{st}}$ band. Thereby, in fulfillment of energy-momentum conservation, an energy per particle of approximately the band gap between the $1^{\textrm{st}}$ and $2^{\textrm{nd}}$ bands is deposited into motion along the $z$-axis. Starting with a balanced spin mixture, it is reasonable to assume that the same dynamical evolution holds for both spin components. In absence of binary collisions, we may hence describe each spin component by the same equations used for the spin-polarized case (Eq.~\ref{eq:relax1}) with according particle numbers $N_1$ and $N_2$ representing the populations of either spin component in the $1^{\textrm{st}}$ and $2^{\textrm{nd}}$ bands. 

In an extended minimal relaxation model, including binary collision transfer between the $2^{\textrm{nd}}$ and $1^{\textrm{st}}$ band, we have to consider an additional class of atoms with population $Z_1$ belonging to the $1^{\textrm{st}}$ band but possessing additional excitation along the $z$-axis with an energy similar to the band gap between the $1^{\textrm{st}}$ and $2^{\textrm{nd}}$ band. Similarly as for the case of $N_2$, also $Z_1$ is subject to a decrease by heating towards the lost atom population $N_L$ at a rate $f_{L}$. The decrease of $N_2$ towards $Z_1$ is modeled by a two-body collision term $g\,N_{2}^2$, with $g \equiv \beta / (2 V_{N_2,\textrm{eff}})$, where $\beta$ denotes the two-body collision parameter and $V_{N_2,\textrm{eff}}$ is the effective Volume of the sample in the state $|N_2\rangle$ \cite{Wei:99, Han:06}. Reversely, the decrease of $Z_1$ towards $N_2$ is given by a two-body collision term $f\,Z_{1}^2$, with $f \equiv \beta / (2 V_{Z_1,\textrm{eff}})$, where $V_{Z_1,\textrm{eff}}$ is the effective Volume of the sample in the state $|Z_1\rangle$. We expect $V_{Z_1,\textrm{eff}} > V_{N_1,\textrm{eff}}$ and hence $f < g$. This relaxation model is sketched in Fig.~\ref{fig3}(c) with the equations 
\begin{eqnarray}
\label{eq:relax2}
\dot{N_1}  &=& g_{12} (N_2 - N_1), \nonumber \\
\dot{Z_1}  &=& g N_{2}^2 - f Z_{1}^2 - f_L Z_1,  \\
\dot{N_2}  &=& f Z_{1}^2 - g N_{2}^2 + g_{12} ( N_1 - N_2)- g_L N_2 \,. \nonumber
\end{eqnarray}
In Fig.~\ref{fig3}(d), initially most atoms are loaded into the $2^{\textrm{nd}}$ band, as illustrated by the band mapping image in Fig.~\ref{fig3}(c), showing the initial distribution of atoms across the BZs at $t=10\,$ms with most atoms seen in the $2^{\textrm{nd}}$ BZ. The total populations detected in the first (red symbols) and second (blue symbols) bands, $N_1 + Z_1$ and $N_2$, respectively, are plotted versus the hold time. The solid lines are obtained by using the heating rates $g_{12} =0.0456$ and $g_{L}=0.0432$, found for the spin-polarized case, and by determination of $g = (6.472 \pm 0.234)\times 10^{-6}$, $f = (1.936 \pm 0.173)\times 10^{-6}$ and $f_{L}=0.0167\pm0.0015$ via simultaneously fitting the model of Eq.~\ref{eq:relax2} to both data sets in Fig.~\ref{fig3}(d). Note that a significantly faster, clearly non-exponential decay of $N_2$ is observed as compared to Fig.~\ref{fig3}(b). We may roughly estimate $\beta \approx \eta \,\sigma \,\bar{v}$, where $\sigma = 4 \pi \, a^2$ is the free-space scattering cross section with the scattering length $a = 176\,a_0$, $\bar{v} = \sqrt{8 k_B T/ m \pi}$ is the mean thermal velocity for the temperature $T = 30\,$nK, and $m$ is the atomic mass of potassium. The factor $\eta$ accounts for a transition matrix element involving the initial and final wave functions before and after the collision in the lattice potential. In previous experiments with bosons, small $\eta \ll 1$ have been found to give rise to long lifetimes of higher bands \cite{Koc:16}. With $V_{N_2,\textrm{eff}} \equiv N_{2}^{2} / \int dr^3 \,n_{2}^2(r) \approx  10^{-8}\, \textrm{cm}^{3}$, where $n_{2}(r)$ is roughly approximated by the density profile in the dipole trap, one obtains $g \approx 6 \times 10^{-4}\, \eta$. By comparison with the value determined in the context of Eq.~\ref{eq:relax2} one finds $\eta \approx 10^{-2}$.

In summary, selected excited Bloch bands of an optical square lattice have been loaded with a quantum degenerate Fermi gas with a single or two balanced spin components. In the former case, extreme band lifetimes (> 10 s) are observed as a result of the suppression of collisions due to Pauli's principle. For spin mixtures the lifetime is still on the order of a second although limited by binary collisions between different spin components. The techniques demonstrated here form the basis for simulating fermionic superfluidity in orbital optical lattices. Similar techniques also apply for a wide range of other lattice geometries, including the hexagonal boron-nitride lattice \cite{Wei:16} or the Lieb lattice \cite{Lib:16}, known from cuprate high-temperature superconductors.

\begin{acknowledgments}
We acknowledge support from the Deutsche Forschungsgemeinschaft (DFG) through the collaborative research center SFB 925 (Project No. 170620586, C1). M.H. was partially supported by the Cluster of Excellence CUI: Advanced Imaging of Matter of the Deutsche Forschungsgemeinschaft (DFG) - EXC 2056 - project ID 390715994. We thank Lianghui Huang for useful discussions.   
\end{acknowledgments}

\section*{Supplemental Material}

\subsection{General considerations}
The design and formation of well controlled laser-induced periodic lattice potentials is a central technical challenge in experiments with quantum gases in optical lattices \cite{Gry:01, Lew:07}. If the interest is constrained to physics in the lowest Bloch band of a conventional lattice scenario providing a single class of potential wells, the only relevant tunable parameter is the overall lattice depth. The precise control of discrete symmetries, e.g. with respect to rotation, is usually not required, since only a single ground state is provided on each lattice site. If multipartite unit cells or higher bands are of interest, the situation changes. Lattices composed of local potential wells with tunable relative potential offsets give rise to the possibility of adjusting quantum degeneracies, which requires precise control of the discrete symmetries. Similarly, if higher bands are involved, degenerate states can arise due to orbital degrees of freedom, which likewise requires precision control of the discrete symmetries. The present supplementary material discusses the case of a bipartite square lattice potential, used to confine atoms in higher Boch bands, such that precision control of a multitude of parameters is required. For example, in the second band, two orthogonal nearly degenerate local $p$ orbitals arise. The precise adjustment of their energy separation requires best possible control of the discrete rotational symmetry of the lattice potential. Small imbalances in the intensities or mutual relative angles of the superimposed laser beams can tune this energy separation and hence significantly effect the physical properties of the lattice, if atoms are loaded.

The extra control required here is achieved in a lattice formed in a Michelson-Sagnac interferometer. In Sec. \ref{LatticeSetup}, after the lattice set-up is introduced, the calibration of the relevant control parameters is discussed. In Sec. \ref{XpointDegeneracy}, we present measurements that show that the control of the lattice parameters permits us to shape the energy landscape of the second Bloch band with well below nanoklevin precision. In particular, we can adjust perfect degeneracy of the two inequivalent local energy minima of this band arising at different edges of the first Brillouin zone. Our calibration procedure thus enables us to adjust the lattice potential close to ideal C4 rotation symmetry. In Sec. \ref{FermiGasPrep}, the preparation of spin-polarized samples or balanced mixtures of quantum degenerate fermionic potassium atoms is described. In Sec. \ref{Excitation}, the excitation of such samples into higher Bloch bands of the optical lattice is discussed. Finally, Sec. \ref{Tunneling} shows an examples, how the fast control of the lattice parameters can be used to characterize the tunneling dynamics between within a selected band, and to identify the associated timescales. 

\begin{figure*}[ht]
\includegraphics[width=1\textwidth]{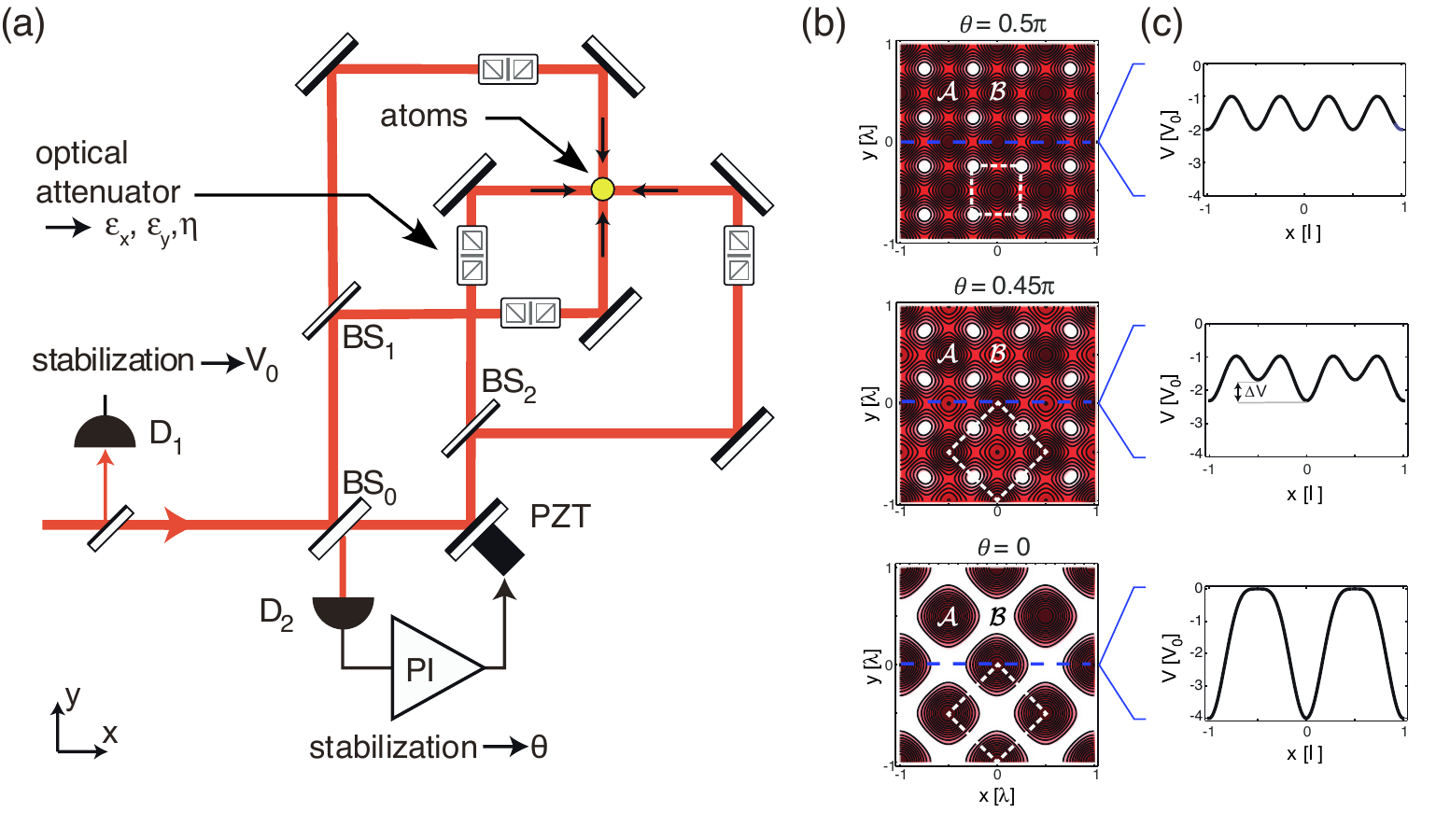}
\caption{\textbf{(a)} Sketch of the optical lattice setup in a Michelson-Sagnac interferometer according to Eq.~(\ref{eq:potential}). The signal at photo detector D$_1$ is used to actively stabilize the power of the beam coupled to the interferometer and hence $V_0$. The interference signal recorded at the photo detector D$_2$ is used to stabilize the time phase difference $\theta$ and therefore the potential difference $\Delta V$ between neighboring lattice sites (cf. (b)). Optical attenuators in the Sagnac loops permit control of the lattice parameters $\epsilon_{x}, \epsilon_{y}$ and $\eta$. \textbf{(b)} Optical potentials for three different time phases $\theta \in \{0.5,0.45,0\} \times \pi$. The white dashed squares show the respective unit cells. The details on the right show the potential along the dashed blue lines.}
\label{fig_S1}
\end{figure*}

\subsection{Lattice setup in a Michelson-Sagnac interferometer}
\label{LatticeSetup}
The optical lattice is realized in a Michelson-Sagnac interferometer, i.e., a conventional Michelson interferometer (as in Ref.~\cite{Hem:92}) with the light in the two branches reflected by Sagnac loops instead of conventional mirrors (cf. Fig.~\ref{fig_S1} (a)). The two loops provide a crossing point, where two nearly orthogonal optical standing waves arise in the $xy$-plane, one defined to be parallel to the $x$-axis and one enclosing a small angle $\xi$ with the $y$-axis, which is unavoidable in experiments. The linear polarizations are parallel to the $z$-axis, such that both standing waves interfere. The induced light shift potential is 
\begin{eqnarray}
\label{eq:potential}
&V(x,y)& = \frac{1}{4}\,V_{0} \bigg|(e^{ikx} + \epsilon_{x} e^{-ikx}) \\ \nonumber
 &+& e^{i\theta} \eta\, (e^{ik(\cos(\xi)y +\sin(\xi)x)} + \epsilon_{y} e^{-ik(\cos(\xi)y +\sin(\xi)x)})\bigg|^2 
\end{eqnarray}
Here, $k=2\pi/\lambda$ and $\lambda=\SI{1064}{\nano\m}$ is the wavelength of the lattice beams. The parameters $\epsilon_{x}$, $\epsilon_{y}$ and $\eta$ with positive values typically close to unity, account for the different intensities of the four superimposed laser beams due to different powers or beam sizes. The angle $\theta$ denotes the time phase difference between the resulting standing waves of the two lattice axes. The parameters $\epsilon_{x}$, $\epsilon_{y}$, $\eta$ can be individually adjusted by means of the four electrically driven optical attenuators shown in Fig.~\ref{fig_S1} (a). The phase angle $\theta$, determined by the difference of the optical path length between the beam splitters BS$_0$ and BS$_1$ and that between the beam splitters BS$_0$ and BS$_2$, and by the optical path length difference of the two loops, is actively stabilized with a precision of $10^{-3} \pi$ by locking the interference fringe signal recorded at detector D2 to a constant value. An additional weak frequency component at $\lambda=\SI{1083}{\nano\m}$ is coupled to the interferometer for this purpose, which can be readily discriminated from the $1064\,$nm light and does not provide a notable potential for the atoms. As illustrated in Fig.~\ref{fig_S1} (b), the $1064\,$nm potential provides two classes of wells, denoted $\mathcal{A}$ and $\mathcal{B}$, with a relative potential offset $\Delta V$ adjusted by the choice of $\theta$ according to $\Delta V/V_0 \approx \cos(\theta)\, \eta \,(1+\epsilon_{x})(1+\epsilon_{y})$. The use of Sagnac loops permits one to readily adjust the parameters $\epsilon_{x}$, $\epsilon_{y}$ and $\eta$ to approach unity with high precision. There are six choices of different pairs of lattice beams, which interfere to form a 1D lattice structure. For each of these lattices one determines the well depth by means of measuring the resonance frequency for parametric excitation. Few iterations allow one to approximate $\epsilon_{x} = \epsilon_{y} = \eta = 1$ to better than a percent. If in addition $\xi = 0$ can be adjusted, the lattice acquires C4 rotation symmetry. Adjustment of $\xi$ to zero, however, requires a complex protocol involving manual intervention. A straight forward observable, only based on the detection of light, that lets one determine the value of $\xi$, is not available. In Sec. \ref{XpointDegeneracy}, we discuss how atoms Bose-condensed in excited bands can be used to obtain precise information on the value of $\xi$.

\begin{figure*}[ht]
\includegraphics[width=1\textwidth]{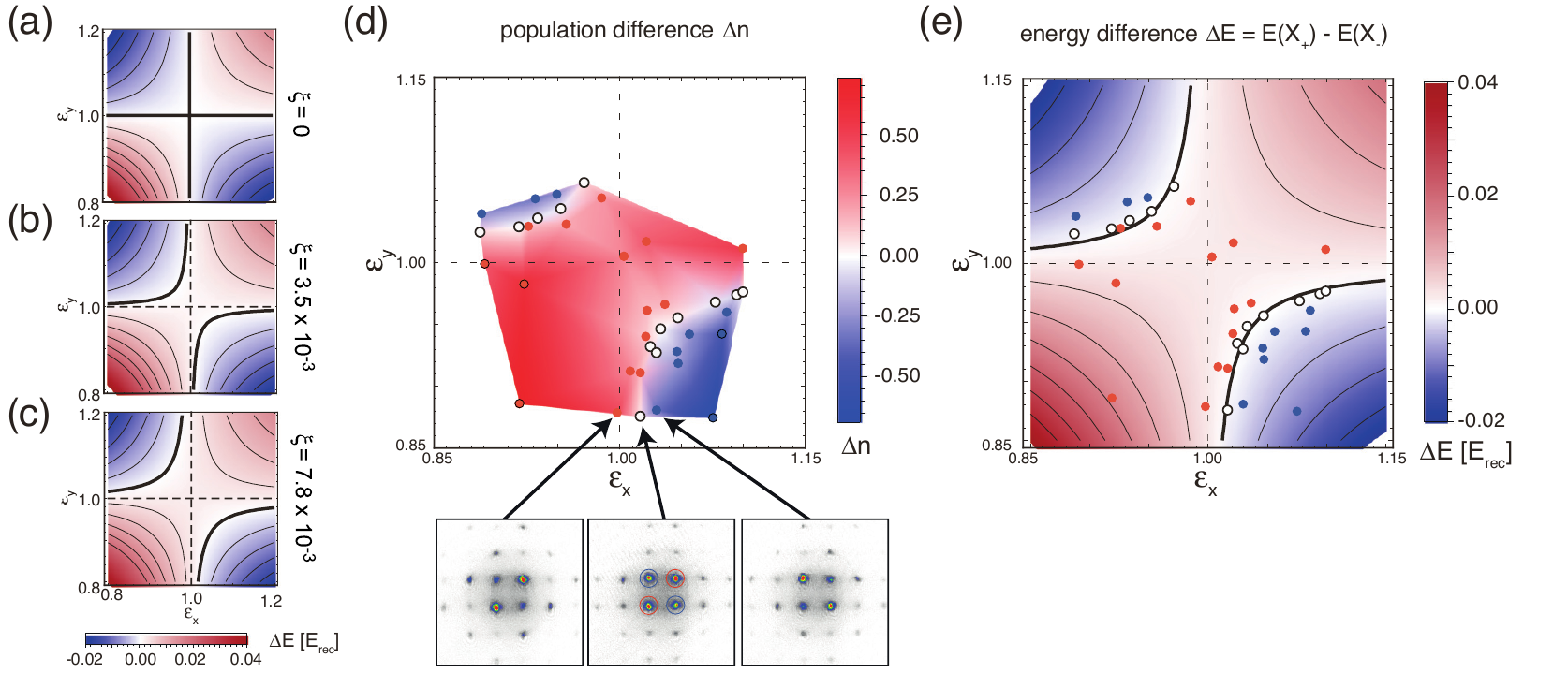}
\caption{\textbf{(a),(b),(c)} Single particle band calculations of $\Delta E$ versus $\epsilon_{x}$ and $\epsilon_{y}$ for $\xi = 0$, $\xi = 3.5\times 10^{-3}$ and $\xi = 8.7\times 10^{-3}$. \textbf{(d)} Experimental determination of $\Delta n$ versus $\epsilon_{x}$ and $\epsilon_{y}$. \textbf{(e)} Calculation of $\Delta E$ for $\xi = 4.5\times 10^{-3}$ superimposed with the data in (d). For the entire figure, $\eta=1$, $\theta = 0.54 \pi$ and $V_0= 7 E_{\textrm{rec}}$.}
\label{fig_S2}
\end{figure*}

\subsection{Controlling energy momentum dispersion of the second Bloch band}
\label{XpointDegeneracy}
The detailed control of the lattice parameters $\epsilon_{x}$, $\epsilon_{y}$ and $\eta$, offered by the Michelson-Sagnac interferometer design, allows us to precisely engineer the band structure. By loading a Bose-Einstein condensate to the second Bloch band, we obtain a probe that lets us observe and hence adjust orbital degeneracies between $p_x$ and $p_y$ orbitals in order to approach approximate C4 rotation symmetry of the lattice potential, via adjustment of $\epsilon_{x} \approx \epsilon_{y} \approx \eta \approx 1$ and $\xi \approx 0$. The second band provides two inequivalent high symmetry points ($X_{+}$ and $X_{-}$) at the edge of the first Brillouin zone, where the energy momentum dispersion provides local minima in quasi-momentum space, denoted $E(X_{+})$ and $E(X_{-})$, respectively (cf. Fig.9(a) in Ref.~\cite{Koc:16}). The energy difference $\Delta E \equiv E(X_{+}) - E(X_{-})$ (derived from exact band calculations for the potential in Eq.~(\ref{eq:potential})) is plotted versus $\epsilon_{x}$ and $\epsilon_{y}$ with fixed $\eta=1$ in Figs.~\ref{fig_S2} (a),(b),(c) for three values of the angle $\xi$, i.e. $\xi = 0$, $\xi = 3.5\times 10^{-3}$, $\xi = 8.7\times 10^{-3}$, respectively. As is seen in Fig.~\ref{fig_S2}(a), for $\xi = 0$, degeneracy of the $X$-points (i.e., $\Delta E = 0$) arises on the $\epsilon_{x}$ and $\epsilon_{y}$-axes. Only in the origin, C4 symmetry prevails, showing that $X$-point degeneracy does not require C4 symmetry. If $\xi$ even slightly deviates from zero (cf. Figs.~\ref{fig_S2}(b),(c)), C4 symmetry is not available for any values of $\epsilon_{x}$, $\epsilon_{y}$, while $X$-point degeneracy is still available on hyperbolas, highlighted by thick black lines. Observation of $\Delta E(\epsilon_{x},\epsilon_{y})$ can be used as a monitor to adjust C4 symmetry.

In order to map out $\Delta E(\epsilon_{y},\epsilon_{y})$, bosonic rubidium atoms ($^{87}$Rb) can be employed. Using the methods described in Ref.~\cite{Koc:16}, $^{87}$Rb atoms can be loaded into the second band, where a Bose-Einstein condensate is formed with condensate fractions at both $X$-points $n(X_{+})$ and $n(X_{-})$. At the lower edge of Fig.~\ref{fig_S2}(d), three exemplary momentum spectra are shown recorded at positions in the $(\epsilon_{x},\epsilon_{y})$-plane indicated by arrows. Such spectra are obtained by switching off the lattice and trap potentials and allowing for a ballistic flight after which an absorption image is recorded (for details see Ref.~\cite{Koc:16}). The condensate fractions $n(X_{+})$ and $n(X_{-})$ are determined by counting the atoms within the disk-shaped regions enclosed by red and blue circles, respectively. The relative condensate fraction difference $\Delta n=\frac{n(X_{+})-n(X_{-})}{n(X_{+})+n(X_{-})}$ is plotted versus $\epsilon_{x}$ and $\epsilon_{y}$, using the color scale shown at the right edge of Fig.~\ref{fig_S2}(d). The disk-shaped symbols mark the positions in the $(\epsilon_{x},\epsilon_{y})$-plane, where a measurement was performed. Measurements with $|\Delta n | < 0.1$ are indicated by white disks. The colored region is obtained by extrapolating between these measurements. The quantity $\Delta n$ is directly proportional to $\Delta E$ and hence permits a comparison with the theory in Figs.~\ref{fig_S2}(a),(b),(c). The optimal agreement arises for $\xi = 4.5\times 10^{-3}$, corresponding to $0.26^\circ$, which is shown in Fig.~\ref{fig_S2}(e). The disk-shaped symbols are the same shown in Fig.~\ref{fig_S2}(d). The white $\Delta n \approx 0$ disks are well described by the calculated black $\Delta E = 0$ hyperbolas. Fig.~\ref{fig_S2}(e) shows that a deviation of $\xi$ from zero by $0.26^\circ$ amounts to a change of $\Delta E$ of about $10^{-3}\,E_{\textrm{rec}}$ corresponding to $0.1\,$nK, which can be compensated by changes of $\epsilon_{x}$ and $\epsilon_{y}$ on the order of a few percent.

\subsection{Preparation of $^{40}$K Fermi gas}
\label{FermiGasPrep}
A degenerate Fermi gas of potassium atoms ($^{40}$K) in the $\ket{F=9/2, m_F=9/2}$ hyperfine state is formed in a conventional two-species quantum gas machine, which permits to simultaneously process bosonic rubidium atoms ($^{87}$Rb), serving as a coolant. After loading and pre-cooling both species in a combined magneto optical trap and subsequent cooling stages via grey molasses \cite{Fer:12, Ros:18}, the atoms are magnetically transferred into a magnetic quadrupole trap. Here, $^{40}$K $\ket{F=9/2, m_F=9/2}$ atoms are sympathetically cooled by $\ket{F=2, m_F=2}$ $^{87}$Rb atoms, which are cooled via radio-frequency (RF) evaporation. Before the temperature reaches values, where spin-flip losses in the trap center set in, an optical dipole trap, made of two crossed laser beams, propagating in the $xy$-plane with a wavelength of $1064\,$nm, is ramped up, while the magnetic gradient of the quadrupole trap is reduced to 7 G/cm, such that gravity is compensated for $^{40}$K. Finally, $^{40}$K is further cooled sympathetically via optical evaporation of $^{87}$Rb by reducing the depth of the dipole trap. Due to the different masses of $^{87}$Rb and $^{40}$K, the magnetic field compensates only the gravitation for $^{40}$K such that mainly $^{87}$Rb is evaporated out of the dipole trap, while the loss of $^{40}$K remains moderate. We end up with a Fermi gas of $10^{5}$ optically trapped $^{40}$K atoms in the $\ket{F=9/2, m_F=9/2}$ state with a temperature $0.12\,T/T_{F}$, where $T_{F}$ denotes the Fermi temperature. At this stage, the remaining $^{87}$Rb atoms can be optionally removed by applying a short pulse of resonant laser radiation.

In order to produce $^{40}$K spin mixtures in the $\ket{F=9/2, m_F=-9/2}$ and $\ket{F=9/2, m_F=-7/2}$ states, the following protocol is applied. We begin with a cold $^{40}$K sample in the $\ket{F=9/2, m_F=9/2}$ state in the dipole trap with a homogeneous $1\,$G bias magnetic field in the $z$-direction applied. The homogeneous magnetic field is ramped up in $10\,$ms to a value of $27.75\,$G and held at this value for $10\,$ms. Next, RF radiation at $10.44\,$MHz  is applied by an antenna with a power giving rise to a $0.35\,$G peak magnetic field at the position of the atoms. Next, the magnetic field is ramped up in $5\,$ms from $27.75\,$G to $36.8\,$G, which gives rise to an adiabatic passage from $\ket{F=9/2, m_F=9/2}$ to $\ket{F=9/2, m_F=-9/2}$. In order push $50\%$ of these atoms to $\ket{F=9/2, m_F=-7/2}$, a near-resonant Rabi $\pi/2$ pulse is used. To this end, the RF is tuned down in power by $60\,$dB and the magnetic field is ramped to $35.6\,$G in $2\,$ms, i.e., resonance for the transition between the bare Zeeman states $\ket{F=9/2, m_F=-7/2}$ and $\ket{F=9/2, m_F=-9/2}$ is established. Next, the RF power is increased to the previous level again for $2.2\,\mu$s, which yields approximately a 1:1 population of $\ket{F=9/2, m_F=-7/2}$ and $\ket{F=9/2, m_F=-9/2}$ to better than a few percent. The resulting spin populations are controlled with a Stern-Gerlach method, where subsamples occupying different spin components are spatially separated by an inhomogeneous magnetic field and are counted. The precision of spin preparation is a few percent. 

\begin{figure}[ht]
\includegraphics[width=7.4 cm]{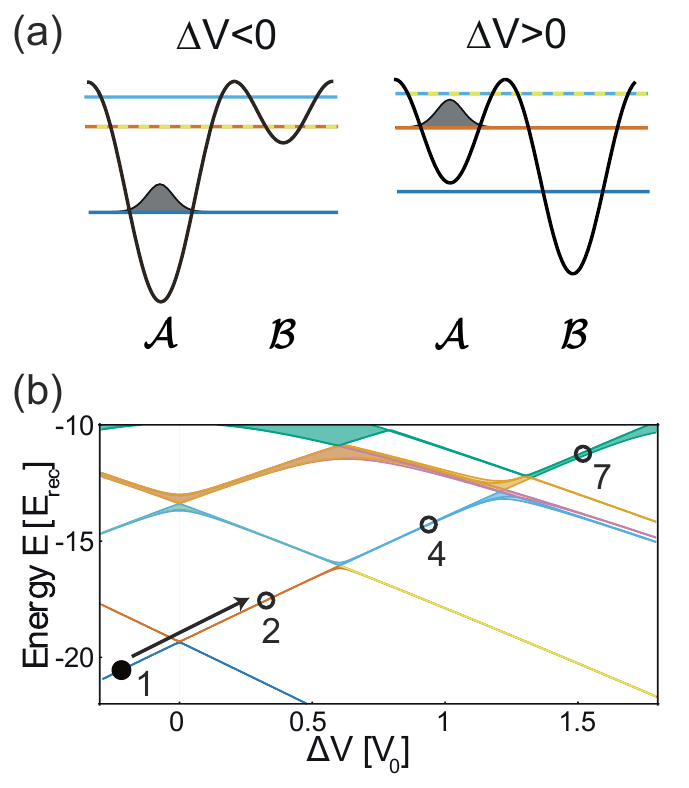}
\caption{(a) Sketch of the excitation protocol. A section of the lattice potential is shown with two adjacent $\mathcal{A}$- and $\mathcal{B}$-wells with the first four Bloch bands indicated by horizontal lines. Two-color dashed lines denote twofold degenerate bands. The predominantly occupied local orbital (Wannier function) is highlighted. See text for details. (b) The lowest seven energy bands are plotted against the potential difference $\Delta V$ for $V_{0}= 12\,E_{\textrm{rec}}$. The black disk indicates the starting point for the excitation protocol corresponding to the left panel in (a). The black circles indicate possible end points of the excitation quench, leading to selective population of the band with the band index indicated below the circle.}
\label{fig_S3}
\end{figure}

\subsection{Excitation of higher bands}
\label{Excitation}
The protocol for preparation of fermionic atoms in selected higher Bloch bands follows the one successfully applied in the case of bosonic atoms, summarized in Ref.~\cite{Koc:16}. The key steps are illustrated in Fig.~\ref{fig_S3}(a). Initially, the atoms (spin-polarized or spin mixtures) are adiabatically loaded into the lattice potential of Eq.~\ref{eq:potential} with $\Delta V = \Delta V_{i} < 0$, such that they exclusively populate the $s$-orbitals of the deeper $\mathcal{A}$-wells and hence belong to the lowest Bloch band. This is achieved by slowly ramping up $V_{0}$ from zero to $12 E_{\textrm{rec}}$ in $150\,$ms, i.e., on a time-scale that is long with respect to the tunneling time with $E_{\textrm{rec}} \equiv \hbar^2 k^2 / 2 m$ denoting the single-photon recoil energy and $m$ the atomic mass. Subsequently, in $100\,\mu$s, which is much faster than the tunneling time, $\Delta V$ is tuned to the final value $\Delta V_{f} > 0$. Hence, the atoms remain trapped in the $\mathcal{A}$-wells, however, elevated with respect to their potential energy such that they belong to an excited Bloch band with a band index adjusted by the choice of $\Delta V_{f}$.

The operating principle of this quench is readily understood via Fig.~\ref{fig_S3}(b), which shows the single-particle band structure plotted versus $\Delta V$. The black disk plotted across the first band at a negative value of $\Delta V_{i}$ indicates the starting point of the quench, while the open circles indicate possible final positive values of $\Delta V_{f}$ chosen such that the $2^{\textrm{nd}}$, $4^{\textrm{th}}$ and $7^{\textrm{th}}$ bands are populated, respectively. The non-adiabaticity of the quench ensures that intersecting band crossings are skipped. Efficient selective population of a single excited band becomes possible, if $\Delta V_{f}$ can be adjusted such that the energy gaps between the addressed band and its neighbouring bands are sufficiently large. In our two-dimensional (2D) bipartite lattice, this can be achieved for band indices of the form $n = 1 + N(N+1)/2$ with $N \in \{1,2,3,...\}$. This is understood by recalling that in a single harmonic 2D potential well, which approximately models the deep wells in our lattice, the degeneracy of the eigenstates equals their principle quantum number.

\subsection{Detection methods}
\label{Detect}
The populations of the Bloch bands are observed by means of a standard technique referred to as band mapping, which proceeds according to the following protocol. First, the lattice potential is ramped down exponentially in $3\,$ms. This time is sufficiently long such that band populations are preserved if no band crossing occurs as the lattice potential is lowered. This is the case for the first $5$ bands in our lattice, while the 6th and 7th bands in fact undergo such a crossing. Thus, the population of the nth band is transferred to the nth Brillouin zone (BZ) for $n\in\{1,2,3,4,5\}$, while the populations of the 6th and 7th bands are transferred to both, the 6th and 7th BZ. After a subsequent $19\,$ms ballistic expansion, an absorption image of the atomic distribution is recorded, which displays an image of momentum space, showing a map of the populations across all BZs. Regular momentum spectra are obtained by replacing the adiabatic decrease of the lattice potential by an instantaneous shut-off in less than a microsecond. Subsequently, the same ballistic expansion during $19\,$ms is applied, followed by the recording of an absorption image. 

\section{Tunneling dynamics}
\label{Tunneling}
\begin{figure}[ht]
\includegraphics[width=8.5 cm]{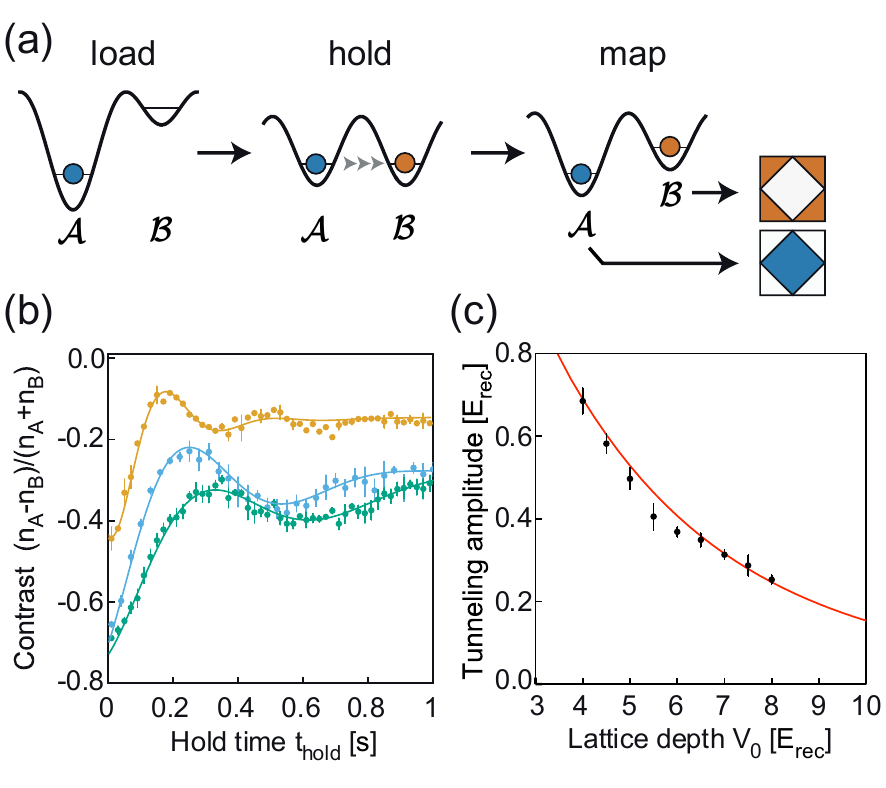}
\caption{(a) Experimental protocol for observing tunneling dynamics: Atoms are initially loaded only to the $\mathcal{A}$-wells adjusting a negative $\Delta V<0$. Subsequently the lattice is quenched to $\Delta V=0$, such that $\mathcal{A}$- and $\mathcal{B}$-wells have the same depth. After a variable hold time, the lattice is quenched back to a negative $\Delta V<0$ and the number of atoms in the $\mathcal{A}$- and $\mathcal{B}$-wells are determined via band mapping. (b) Observed contrast between $\mathcal{A}$- and $\mathcal{B}$-wells plotted versus the holding time (colored disks) for $V_0 / E_{\textrm{rec}}=$ 4 (orange), 6 (blue), 7 (green). The error bars reflect the statistics for on average 8 measurements per data point. The line graphs show fits by exponentially decaying harmonic oscillations. (c) Oscillation frequencies extracted according to (b) plotted against the lattice depth. The error bars show the standard deviations of fits analogous to those in (b). The red line graph shows two times the bandwidth of the first band derived by an exact band calculation.}
\label{fig_S4}
\end{figure}

The fast control of the potential difference $\Delta V$ can be used to investigate the tunneling dynamics between the $\mathcal{A}$- and $\mathcal{B}$-wells within a selected band, and to identify the associated timescales. For the example of the first band, the protocol shown in Fig.~\ref{fig_S4}(a) is applied. Initially, the first band of the lattice is adiabatically loaded at a large negative value of $\Delta V$ such that the atoms predominantly populate the deep $\mathcal{A}$-wells. Next, $\Delta V$ is rapidly ($< 200\,\mu$s) tuned to zero, such that $\mathcal{A}$- and $\mathcal{B}$-wells exhibit the same depth and resonant tunneling can occur, resulting in the population of the previously empty $\mathcal{B}$-wells. After a varying hold time $t_{hold}$ the lattice is quenched a second time back to a negative $\Delta V<0$, such that the atoms in the $\mathcal{A}$-wells are projected to the first band, while those in the $\mathcal{B}$-wells are projected to the second band. Hence, band mapping (see main text) detects $\mathcal{A}$-atoms in the first and $\mathcal{B}$-atoms in the second BZ, such that the number of atoms $n_A$, $n_B$ in the $\mathcal{A}$ and $\mathcal{B}$ wells can be readily counted.

The tunneling dynamics can be thus studied by varying the hold time. In Fig.~\ref{fig_S4}(b), the observed contrast $(n_A - n_B)/(n_A + n_B)$ of atoms in $\mathcal{A}$- and $\mathcal{B}$-wells is plotted versus the hold time for three exemplary lattice depths $V_0 / E_{\textrm{rec}}=$ 4 (orange), 6 (blue), 7 (green). One recognizes a damped oscillation due to Rabi dynamics between $\mathcal{A}$- and $\mathcal{B}$-wells at a frequency given by the tunneling amplitude. These observations can be qualitatively modeled by a simple two-level model. The observed damping is attributed to decoherence resulting from the finite system size and the trap potential, which leads to different tunneling amplitudes at different positions in the lattice. The oscillation frequencies and damping times can be readily extracted by fits with an exponentially decaying harmonic oscillation. In Fig.~\ref{fig_S4}(c) (black disks) the tunneling amplitudes thus obtained are plotted against the lattice depth $V_0 / E_{\textrm{rec}}$. The expectation that in the tight binding regime, i.e. for sufficient lattice depth, the tunneling amplitudes are directly proportional to the band width is confirmed by the red line graph, which plots two times the band width of the first band obtained by an exact band calculation.

\end{document}